
\magnification = 1200
\baselineskip = 14pt
\nopagenumbers
\def\titolo#1{\vglue 2truecm \centerline {\bf #1} \medskip}
\def\autori#1{\centerline {\it #1} \bigskip \bigskip \medskip}
\def\referenze#1{\vglue 3 truecm {\bf References} \bigskip #1}

\def\undertext#1{$\underline{#1}$}

\vglue 3truecm

\centerline{\bf ITALIAN WORKSHOP ON QUANTUM GROUPS}
\centerline{\bf Florence, February 3 to 6, 1993}

\bigskip
\bigskip

\centerline{edited by {\it E. Celeghini and M. Tarlini}}
\centerline{\it Dipartimento di Fisica, INFN Sezione di Firenze}

\vskip 7truecm

\centerline{\bf Foreword}

\medskip

The first ``{\it Convegno Informale su Quantum Groups}'' was held in
Florence from February 3 to 6, 1993. This {\it Convegno} was conceived
as an informal meeting to bring together all the italian people
working in the field of quantum groups and related topics.
We are very happy indeed that about 30 theoretical physicists decided
to take part presenting many aspects of this interesting and live
subject of research.
We thank all the participants for the stimulating and nice atmosphere
that has characterized the meeting.

This paper has the intent to give
a quick review in english of the contributions and related references.

We think useful to include the complete addresses and coordinate data of
the participants. It is our intention to diffuse these proceedings
 by e--mail trough electronic data banks.

\bigskip
\bigskip

\halign{\hglue 8truecm #\ &\quad#\cr
Enrico Celeghini&Marco Tarlini\cr
celeghini@fi.infn.it&tarlini@fi.infn.it\cr}

\vfill\break

\footline={\hss\tenrm\folio\hss} \pageno=2

\vglue 2truecm

\centerline{\bf Contents}

\bigskip
\bigskip

\noindent
A.Montorsi and M.Rasetti,~~{\it Q-symmetries of the Hubbard model
with phonons}\hfill 3\break
A.Liguori and  M.Mintchev,~~{\it On Zamolodchikov's equation}\hfill
5\break
L.Castellani,~~{\it Gauge theories of quantum unitary groups}\hfill
7\break
E.Celeghini, M.Rasetti and G.Vitiello,~~{\it Q-derivatives, coherent
states and squeezing}\hfill 10\break
L.Lusanna,~~{\it
Symplectic approach to relativistic localization:
Dirac-Yukawa\hfill\break \hglue 2.35truecm ultraviolet cutoff or
classical basis
of the Manin quantum plane?}\hfill 11\break
G.Fiore,~~{\it Realization of $U_q(so(3))$ within $Diff({\bf
R}_q^3)$}\hfill 13\break
F.Bonechi, E.Celeghini, R.Giachetti, E.Sorace and M.Tarlini,~~{\it
Quantum groups and\hfill\break \hglue 11.6truecm
lattice physics}\hfill 15\break
L.Bonora,~~{\it Toda field theories and quantum groups}\hfill 18\break
A.Sciarrino,~~{\it ``Embedding'' of q-algebras}\hfill 20\break
A.Lerda and S.Sciuto,~~{\it Anyons and Quantum Groups}\hfill 22\break
P.Truini,~~{\it From classical to quantum: the problem of
universality}\hfill 23\break
L.Martina, O.Pashaev and G.Soliani,~~{\it Anyons in planar
ferromagnets}\hfill  25\break
L.Dabrowski,~~{\it Positive energy unitary representations
of the conformal q-algebra}\hfill 27\break
P.Cotta--Ramusino, L.Lambe and M.Rinaldi,~{\it Construction of
quantum groups and\hfill\break \hglue 8.5truecm
the Y.B.E. with spectral parameter}\hfill 29\break
E.Guadagnini,~~{\it Topological field theory and invariants
of three-manifolds} \hfill  31\break
A.Nowicki, E.Sorace and M.Tarlini,~~{\it The quantum Dirac
equation associated\hfill\break \hglue 6.9truecm  to the
$\kappa$--Poincar\'e}\hfill   33\break
R.Floreanini and L.Vinet,~~{\it Quantum algebras and basic
hypergeometric functions }\hfill  35\break
M.Carfora, M.Martellini and A.Marzuoli,~~{\it 4--Dimensional
lattice gravity \hfill\break \hglue 8truecm and 12j-symbols}\hfill 37\break
D.Franco and C.Reina,~~{\it The geometrical meaning of the quantum
correction}\hfill 39\break
\smallskip\noindent
Participants\hfill 40\break

\vfill\break


\def\xj{x_{\bf j}}
\def\pj{p_{\bf j}}

\def\pk{p_{\bf k}}
\def\tjk{t_{{\bf j}, {\bf k}}}

\def\ais{a_{{\bf j}, \sigma}}

\def\ajsd{a_{{\bf k},\sigma}^\dagger}

\def\couli{n_{{\bf j}, \Uparrow} n_{{\bf j}, \Downarrow}}
\def\nup{n_{{\bf j}, \Uparrow}}
\def\ndw{n_{{\bf j}, \Downarrow}}
\def\diracij{\delta_{{\bf j}, {\bf k}}}

\def\CcC{{\hbox{\tenrm C\kern-.45em{\vrule height.67em width0.08em
depth-.04em
\hskip.45em }}}}
\def\RrR{{\hbox{\tenrm I\kern-.17em{R}}}}
\def\ZzZ{{\hbox{\tenrm Z\kern-.31em{Z}}}}
\def\NnN{{\hbox{\tenrm {I\kern-.18em{N}}\kern-.18em{I}}}}
\def\IiI{{\hbox{\tenrm I\kern-.19em{I}}}}

\titolo{ Q-symmetries of the Hubbard model with phonons}

\autori{Arianna Montorsi and \undertext{Mario~Rasetti}}

We show that the addition of a phonon field to the Hubbard
model, although breaking the 'superconductive' $su(2)$ symmetry [1] of the
Hubbard hamiltonian itself, restores it
as a deformed ({\sl quantum group}) $[su(2)]_q$ symmetry, where $q$ is related
to the strength of electron-phonon coupling. Moreover, the chemical potential
(and hence the filling) at which the symmetry is restored turns out to depend
on the same interaction strength. This latter feature suggests the possibility
of having 'superconducting' states with off diagonal long range order at a
filling different from a half, with a non-vanishing projection over the ground
state, thus reproducing the qualitative behavior of high-$T_c$ materials.

Here the phonons are identified with an
ensemble of independent Einstein oscillators with frequency $\omega$ [2],
and may
be thought of as describing nothing but the ions oscillations around the
lattice positions. When one switches on the phonon field, therefore, the
grand-canonical Hubbard hamiltonian is changed into
$$
H=H_{Hub}^{(loc)} + H_{ph} + H_{el-ph}^{(hop)} \quad , \eqno{(1)}
$$
where (in units in which the ion mass $M=1$)
$$
H_{Hub}^{(loc)}= \sum_{{\bf j}}\left ( - \mu (\nup+\ndw ) +
U\, \couli \right)   \; , \;
H_{ph} = {1\over 2} \sum_{\bf j} \left ( p_{\bf j}^2 +\omega^2 x_{\bf j}^2
\right ) \quad , \eqno{(2)}
$$
$\nup$ and $\ndw$ being the number operators of electrons with up and down
spin, $\xj,\, \pj$ the local ion displacement and momentum operators
respectively ($[\xj,\pk]=i \diracij$) commuting with the fermi operators,
and [3,4]
$$
\eqalign{
H_{el-ph}^{(hop)} = &\sum_{<{\bf j},{\bf k}>} \sum_\sigma \left
(\tjk \, \ajsd \ais + {\rm h.c.} \right ) =
= - \lambda \sum_{\bf j} (\nup+\ndw)\,
x_{\bf j}\cr + &t_0 \sum_{<{\bf j},{\bf k}>} \sum_\sigma \biggl
( exp{\left \{ (-)^{|{\bf j}|} \zeta \left (x_{\bf j} - x_{\bf k}
\right )\right \}}\, exp{\left \{ \kappa \left (p_{\bf j} - p_{\bf k} \right )
\right \}}\, \ajsd \ais + {\rm h.c.} \biggr )\cr }\; \eqno{(3)} .
$$
with $\tjk$ giving the hopping amplitude due to the overlap of the
electron orbitals centered at the displaced ion sites,
$$
\tjk = \int d{\bf r} \phi^*({\bf r}-R_{\bf j}-x_{\bf j})\left [ -{1\over
2}\triangle + V({\bf r}) \right ] \phi ({\bf r}-R_{\bf l}- x_{\bf l}) \quad ,
\eqno{(4)}
$$
and $\lambda$ being the strength of the local electron-phonon coupling
originated by $V({\bf r})$ [3].

The q-deformed 'superconductive' $[su(2)]_q$ symmetry algebra of
hamiltonian (1) is generated by
$$\eqalign{
\hat K^{(z)} &= \sum_{\bf j} \IiI \otimes \cdots \IiI \otimes K^{(z)} \otimes
\IiI \cdots \otimes \IiI = \sum_{\bf j} K_{\bf j}^{(z)}\cr
\hat K^{(+)} &= \sum_{\bf j} {\rm e}^{i {\bf\pi}\cdot{\bf j}}
{\rm e}^{- \alpha K^{(z)}} \otimes \cdots {\rm e}^{- \alpha K^{(z)}} \otimes
{\rm e}^{- \alpha K^{(z)}} \otimes K^{(+)} \otimes {\rm e}^{\alpha^* K^{(z)}}
\otimes \cdots \otimes {\rm e}^{\alpha^* K^{(z)}} \cr
&= \sum_{\bf j} {\rm e}^{i {\bf\pi}\cdot{\bf j}} \prod_{{\bf k}<{\bf j}}
{\rm e}^{- \alpha K_{\bf k}^{(z)}} K_{\bf j}^{(+)} \prod_{{\bf k}>{\bf j}}
{\rm e}^{\alpha^* K_{\bf k}^{(z)}} \quad ; \quad  \hat K^{(-)} =
\left [\hat K^{(+)}\right ]^\dagger \cr} \eqno{(5)}
$$
where
$$
K_{\bf j}^{(+)} = {\rm e}^{i \kappa p_{\bf j}} a_{{\bf j}, \Uparrow}^{\dagger}
a_{{\bf j}, \Downarrow}^{\dagger} \; , \; K_{\bf j}^{(-)} = K_{\bf j}^{(+)\,
\dagger} \; , \; K_{\bf j}^{(z)} = {1\over 2} \left ( \nup + \ndw - 1 \right )
\; \eqno{(6)}
$$
In order for the $[su(2)]_q$-symmetry to hold, it turns out that the parameters
$\mu$, $\kappa$, and $\alpha$ have to be related to the physical parameters
$U$, $\lambda$, $\omega$, and $\zeta$, by the following constraints:
$$
\mu=U-{\lambda^2\over\omega}\quad ,\quad \zeta ={\alpha\over f}\quad ,\quad
\kappa= {\lambda\over\omega}
\quad . \eqno{(7)}
$$

\referenze{
\item{[1]} C.N. Yang, and S.C. Zhang, {\sl Mod. Phys. Lett.} {\bf B4},
759 (1990).
\item{[2]} S. Robaszkiewicz, R. Micnas and J. Ranninger, {\sl Phys. Rev.}
{\bf B36}, 180 (1987).
\item{[3]} H. Fr\"ohlich, {\sl Proc. Roy. Soc. (London)} {\bf A215},
291 (1952).
\item{[4]} A. Montorsi, and M. Rasetti, preprint Politecnico di Torino,
POLFIS-TH-02/93.}

\vfill\break


\titolo{On Zamolodchikov's equation}

\autori{Antonio Liguori and  \undertext{Mihail~Mintchev}}

Recently Carter and Saito [1] discovered a simple but quite remarkable
relationship between the quantum Yang-Baxter (YB) [2] and
the Zamolodchikov tetrahedron (ZT) [3] equations.
This relationship allows in particular to
construct solutions of the ZT equation from solutions of the YB equation.
Let $\cal A$ be an associative algebra
(over ${\bf C}$) with unity ${\bf 1}$. Consider any three elements
$\{ A,\, M,\, B \}$ of ${\cal A}\otimes {\cal A}$, satisfying the YB
$$
A_{12}\, A_{13}\, A_{23} = A_{23}\, A_{13}\, A_{12} \quad ,
\quad \quad
B_{12}\, B_{13}\, B_{23} = B_{23}\, B_{13}\, B_{12} \quad ,
$$
and the mixed equations
$$
M_{12}\, M_{13}\, A_{23} = A_{23}\, M_{13}\, M_{12} \quad ,
\quad \quad
B_{12}\, M_{13}\, M_{23} = M_{23}\, M_{13}\, B_{12} \quad .
$$
We call $\{ A,\, M,\, B \}$ a Carter-Saito (CS)
triplet. It is not difficult to show that any CS triplet gives
rise to a solution of the ZT equation. Indeed, using the decompositions
$$
A \equiv \sum_{i\in I} a_i \otimes a^\prime _i \quad , \quad \quad
B \equiv \sum_{j\in J} b_j \otimes b^\prime _j \quad , \quad \quad
M \equiv \sum_{k\in K} m_k \otimes m^\prime _k \quad ,
$$
one can verify by purely algebraic manipulations that
$$
Z = \sum_{i\in I} \sum_{j\in J} \sum_{k\in K}
\left [a_i\otimes m_k \right ]\otimes
\left [a^\prime _i\otimes b_j \right ]\otimes
\left [m^\prime _k\otimes b^\prime _j \right ]
$$
satisfies the ZT equation
$$
Z_{123}\, Z_{145}\, Z_{246}\, Z_{356} =
Z_{356}\, Z_{246}\, Z_{145}\, Z_{123}
$$
on $[{\cal A} \otimes {\cal A}]^{\otimes 6}$.
Clearly, in order to implement effectively the above method for deriving
solutions of the ZT equation, one should solve the preliminary problem
of constructing CS triplets. This is precisely the problem we address in
this talk (see also [4]).

Given a solution
$$
R = \sum_{i \in I} c_i \otimes c^\prime _i \in {\cal A} \otimes {\cal A}
$$
of the YB equation, we have shown in [5] how to reconstruct a relative
semigroup of spectral parameters ${\cal S}(R)$ belonging to
${\rm End}({\cal A})$. Let
${\cal A}_\ell $ and ${\cal A}_r$
be the subalgebras of $\cal A$ generated by
$\{ {\bf 1},\, c_i : i \in I \}$ and
$\{ {\bf 1},\, c^\prime_i : i \in I \}$ respectively. Define
the subset ${\cal S}_\ell (R) \subset {\rm End }({\cal A}_\ell )$ as follows:
$\alpha \in {\cal S}_\ell (R)$ if and only if
there exists $\beta \in {\rm End} ({\cal A}_r )$ such that
$$
[\alpha \otimes {\rm id} ](R) = [{\rm id} \otimes \beta ](R) \quad .
$$
It is easily seen that ${\cal S}_\ell (R)$ is
actually a semigroup with respect to the
composition of endomorphisms. Introducing
$$
R(\alpha ) \equiv [\alpha \otimes {\rm id}](R) \quad ,
$$
one can also show that $R(\alpha )$ satisfies the spectral YB equation
$$
R_{12}(\alpha_1 )R_{13}(\alpha_1 \alpha_2 )
R_{23}(\alpha_2 ) =
R_{23}(\alpha_2 )R_{13}(\alpha_1 \alpha_2 )
R_{12}(\alpha_1 ) \quad , \eqno(1)
$$
$\alpha_1 \alpha_2$ being the composition of the endomorphisms
$\alpha_1$ and $\alpha_2$.

The idempotent elements of ${\cal S}_\ell (R)$
$$
{\cal I}_\ell (R) \equiv \{\varepsilon \in {\cal S}_\ell (R)
\, :\, \varepsilon^2 = \varepsilon \}
$$
play a distinguished role in the above scheme. In fact,
from eq.(1) it follows that $R(\varepsilon )$
satisfies the YB equation for any $\varepsilon \in {\cal I}_\ell (R) $.
Furthermore, it is an immediate consequence of eq.(1) that
$$
\{ R(\varepsilon_1 ),\, R(\varepsilon_2 \alpha \varepsilon_1), \,
R(\varepsilon_2) \}
$$
is a SC triplet for any $\alpha \in {\cal S}_\ell (R)$ and
$\varepsilon_1 ,\, \varepsilon_2 \in {\cal I}_\ell (R) $.
In this way one obtains a whole
family of CS triplets, naturally generated by a solution of the quantum
YB equation.

Some consequences of the above construction have been explored in [4].
We believe that further investigations in this framework will shed
new light on the relationship between the YB and the ZT equations.

\referenze{
\item{[1]} J.S. Carter and M. Saito, {\it On formulations and solutions of
simplex equations}, Preprint University of Texas at Austin, 1992.
\item{[2]} C.N. Yang, {\sl Phys. Rev. Lett.} {\bf 19}, 1312 (1967);
R. J. Baxter, {\it Exactly Solved Models in Statistical Mechanics}
(Academic Press, New York 1982).
\item{[3]} A.B. Zamolodchikov, {\sl Commun. Math. Phys.} {\bf 79},
489 (1981).
\item{[4]} A. Liguori and M. Mintchev, {\it Some solutions of the
Zamolodchikov tetrahedron equation}, Preprint University of Pisa,
IFUP-TH 13/93.
\item{[5]} A. Liguori and M. Mintchev, {\sl Phys. Lett.} {\bf B275},
371 (1992).}

\vfill\break


\titolo{Gauge theories of quantum unitary groups}

\autori{\undertext{Leonardo~Castellani}}

\def\noi{\noindent}

\def\unmezzo{{1 \over 2}}
\def\epsi{\varepsilon}

\def\de{\delta}

\def\part{\partial}

\def\Lcal{{\cal L}}

\def\R#1#2{ \Lambda^{#1}_{~~#2} }
\def\Rinv#1#2{ (\Lambda^{-1})^{#1}_{~~#2} }

\def\PA#1#2{ {P_A}^{#1}_{~~#2} }

\def\C#1#2{ {\bf C}_{#1}^{~~~#2} }
\def\c#1#2{ C_{#1}^{~~#2} }
\def\q#1{   {{q^{#1} - q^{-#1}} \over {q^{\unmezzo}-q^{-\unmezzo}}}  }

\def\q1{$q \rightarrow 1$}
\def\Fmn{F_{\mu\nu}}
\def\Am{A_{\mu}}
\def\Ama{A_{[\mu}}
\def\An{A_{\nu}}
\def\dm{\part_{\mu}}
\def\dn{\part_{\nu}}
\def\Ana{A_{\nu]}}

\def\dma{\part_{[\mu}}

\def\gij{g_{ij}}
\def\Lcal{{\cal L}}

\def\La{\Lambda}
\def\Lam{\La^{-1}}

In ref.s [1,2] we have proposed a geometric approach to
the construction of
$q$-gauge theories, based on the differential calculus on
$q$-groups developed in
[3-6].These theories
are continuously
connected with ordinary Yang-Mills theories, just as quantum groups are
continuously connected with ordinary Lie groups.
Spacetime is taken to be ordinary
(commutative) spacetime, but the whole discussion holds for
a generic $q$-spacetime. The notations will be as in ref. [6].

The $q$-Lie algebra we obtain from a
bicovariant differential calculus is given
by the $q$-commutations between the quantum generators $T_i$:

$$T_i T_j - \R{kl}{ij} T_k T_l = \C{ij}{k} T_k \eqno(1) $$

\noi The braiding matrix $\Lambda$  and the $q$-structure
constants $\C{ij}{k}$
can be expressed in terms
of the $R$-matrix of the corresponding $q$-group, as
shown in [4] and further discussed in [5,6], and satisfy four conditions:

$$\R{ij}{kl} \R{lm}{sp} \R{ks}{qu}=\R{jm}{kl} \R{ik}{qs} \R{sl}{up}
{}~~~~~~~~{\rm(Yang-Baxter~equation)} \eqno(2a)$$
$$\C{mi}{r} \C{rj}{n} - \R{kl}{ij} \C{mk}{r}
\C{rl}{n} = \C{ij}{k} \C{mk}{n}
{}~~~{\rm (q-Jacobi~identities)} \eqno(2b) $$
$$\C{is}{j} \R{sq}{rl} \R{ir}{pk} + \C{rl}{q} \R{jr}{pk} = \R{jq}{ri}
\R{si}{kl} \C{ps}{r} + \R{jq}{pi} \C{kl}{i} \eqno(2c)$$
$$\R{ir}{mk} \R{ks}{nl} \C{rs}{j}=\R{ij}{kl} \C{mn}{k} \eqno(2d)$$

\noi The last
two conditions are trivial in the limit \q1 ($\R{ij}{kl}
=\de^i_l \de^j_k$).

We start by defining the field strength as

$$\Fmn \equiv \unmezzo (\dm \An - \dn \Am + \Am \An - \An \Am)=
        \dma \Ana - \Ama \Ana \eqno(3)$$

\noi where $\Am \equiv \Am^i T_i$.
The gauge potentials $\An^i$ are taken to satisfy the $q$-commutations:

$$A^i_{[\mu} A^j_{\nu]} =
-{1\over{q^2+q^{-2}}} (\La+\La^{-1})^{ij}_{~~kl}
A^k_{[\mu} A^l_{\nu]}\eqno(4)$$

\noi and simply commute with the quantum
generators $T_i$. The square parentheses
around the $\mu$, $\nu$ indices stand
for ordinary antisymmetrization. The inverse
$\Lam$ of the braiding matrix always exists and is defined
by $\Rinv{ij}{kl} \R{kl}{mn}=\de^i_m \de^j_n$.
The rule (4) is
inspired by the $q$-commutations of
exterior products of left-invariant one-forms
on the quantum groups $U_q(N)$
deduced in ref. [6].
As shown in [2], the field strength (3) can be rewritten as

$$\Fmn^i =\dma \Ana^i + \PA{kl}{mn} \c{kl}{i} \Ama^m \Ana^n  \eqno(5)$$

\noi where $P_A$ is a projector $q$-generalizing the antisymmetrizer,
and $\C{kl}{n} \equiv \c{kl}{n}-\R{ij}{kl} \c{ij}{n}$.
We next define the gauge variations:

$$\de \Am=-\dm \epsi - \Am \epsi + \epsi \Am \eqno(6)$$

\noi with $\epsi \equiv \epsi^i T_i $
and postulate the commutations
$\epsi^i \Am^j = \R{ij}{kl} \Am^k \epsi^l  $.
Under the variations (6) the field strength transforms as
$\de \Fmn = \epsi \Fmn - \Fmn \epsi $. Indeed the
calculation is identical to the usual one, since
both in the definition (3) for $\Fmn$ and in the variations (6) we
have {\sl ordinary} commutators. The $q$-commutativity enters the
game only when we want to factorize the $q$-Lie algebra generators
$T_i$, or
to reorder terms containing $q$-commuting objects like $A,\epsi$.
The ordinary commutator in (6)
leads to a composition law
$(\de_1\de_2-\de_2\de_1) A_{\mu}=-\partial_{\mu} [\epsi_2,\epsi_1]-
[A_{\mu},[\epsi_2,\epsi_1]]$ formally identical
to the classical one.

Using the $A,\epsi$ commutations and eq. (1), the gauge
variations of $\Am^i$ take the familiar form
$\de \Am^i = - \dm \epsi^i - \Am^j \epsi^k \C{jk}{i}$.

By using the bicovariance conditions (2a) and (2d) we can prove that
$\epsi^i F^j = \R{ij}{kl} F^k \epsi^l  $ so that
$ \de \Fmn^i=-\Fmn^j \epsi^k \C{jk}{i}$.
Condition (2c) ensures that the
commutation relations (4)
are preserved under the $q$-gauge
transformations (6), cf. [2].

There is a simple way to obtain the commutations between
$F^i$ and $A^j$, and between $F^i$ and $F^j$, see [2].

Finally, we construct the $q$-lagrangian invariant
under the $U_q(N)$ quantum Lie algebra. We set
$ \Lcal = \Fmn^i \Fmn^j g_{ij}$ where $\gij$, the $q$-analogue
of the Killing metric, is determined by
requiring the invariance of $\Lcal$ under the $q$-gauge transformations
(6).
Under these transformations, the variation of $\Lcal$ is
given by

$$\de \Lcal = -\C{mn}{i} \Fmn^m \epsi^n \Fmn^j \gij - \C{mn}{j}
\Fmn^i \Fmn^m \epsi^n \gij \eqno(7) $$

\noi After reordering the terms as $FF\epsi$ we find that
$\de \Lcal$ vanishes when

$$\C{mn}{i} \R{nj}{rs} \gij + \C{rs}{j} g_{mj} = 0 \eqno(8)$$

\noi This eq. is
not difficult to solve in particular cases. For example,
we have given in ref.[2] the most general
$q$-metric satisfying (8) for the case of
$U_q(2)=[SU(2) \otimes U(1)]_q$.
For conventions, and a detailed study of $U_q(2)$, we refer to [6].

\referenze{
\item{[1]} L. Castellani, {\sl Phys. Lett. } {\bf B292}, 93 (1992).
\item{[2]} L. Castellani, {\it $U_q(N)$ gauge theories}, Torino
preprint DFTT-74/92.
\item{[3]} S.L. Woronowicz, {\sl Publ. RIMS, Kyoto Univ.} Vol. {\bf 23}, 117
(1987); {\sl Commun. Math. Phys.} {\bf 111}, 613 (1987) and
{\sl Commun. Math. Phys.} {\bf 122}, 125 (1989).
\item{[4]} B. Jur\v{c}o, {\sl Lett. Math. Phys.} {\bf 22}, 177 (1991).
\item{[5]} D. Bernard, {\it Quantum Lie
algebras and differential calculus on quantum groups}, Proc. 1990 Yukawa
Int. Seminar, Kyoto; {\sl Phys. Lett.} {\bf B260}, 389 (1991);
U. Carow--Watamura, M. Schlieker, S. Watamura
and W. Weich, {\sl Commun. Math. Phys.} {\bf 142}, 605 (1991);
B. Zumino,
{\it Introduction to the Differential Geometry
of Quantum Groups}, LBL-31432 and UCB-PTH-62/91, notes
of a plenary talk given at the 10-th IAMP Conf., Leipzig (1991);
\item{[6]} P. Aschieri and
L. Castellani, {\it An
introduction to non-commutative
differential geometry on quantum groups}, preprint
CERN-TH.6565/92, DFTT-22/92 (1992), to be publ. in {\sl Int. Jou.
Mod. Phys. {\bf A}.}}

\vfill\break


\titolo{ Q-derivatives, coherent states and squeezing}

\autori{ Enrico Celeghini, Mario Rasetti and \undertext{Giuseppe~Vitiello}}

The q-commutator is discussed in the framework of the Fock-Bargmann
representation and is functionally realized in terms of the commutator of the
z-multiplication operator with the q-derivative (z complex number). We obtain
a weak relation between the q-deformation of the W-H algebra and the
generator of the Glauber coherent states and of the squeezed states.
This study may turn out to be fruitful in relating the quantum algebraic
structures with the theory of the entire analytical functions and of the
theta functions[1].

\bigskip
\bigskip

\referenze{
\item{[1]} E. Celeghini, M. Rasetti and G. Vitiello, in preparation}

\vfill\break


\vglue 2truecm \centerline {\bf
Symplectic approach to relativistic localization:
Dirac-Yukawa}
\centerline{\bf ultraviolet cutoff or classical basis of the Manin
quantum plane?}
\bigskip

\autori{\undertext{Luca~Lusanna}}

The center-of-mass (c.o.m.) and relative variable symplectic basis for
extended relativistic systems (particles, Nambu string, classical fields)
described by first class constraints is described. The action of the
Poincar\'e algebra on the constraint set is such that this set is the disjoint
union of strata, one for each kind of Poincar\'e orbit; in each stratum
the relative variables have the associated Wigner covariance with the
degrees of freedom associated with the relative times playing the role of
gauge variables conjugated to some first class constraint. Instead the
canonical c.o.m. variable is not a four-vector: if one draws in a given
reference frame all the trajectories of this canonical variable (associated
with all possible frames), one obtains a world-tube around the covariant
(but not canonical) Fokker c.o.m. variable. The invariant radius of this
world-tube is an intrinsic classical length determined by the Poincar\'e
Casimirs: it could the classical basis of a deformation parameter for the
definition of the quantum Manin plane; since this length is Casimir dependent
it is not clear which role, if any, it could play in the developments of
the quantum Poincar\'e groups. To make frame independent statements at the
classical level about an extended relativistic system, one cannot localize
its symplectic c.o.m. inside the  world-tube. The standard canonical
quantization with these localization restrictions allows to define an intrinsic
invariant ultraviolet cutoff, when the configuration of the extended system
corresponds to an irreducible Poincar\'e representation with $P^2 > 0$ and
$W^2\not= 0$; it is the Compton wave length multiplied the value of the
rest frame spin. The theorems of Hegerfeldt about the violation of
Einstein causality, when one studies the spreading of localized wave packets
of the Newton-Wigner position operator, imply that the canonical c.o.m.
position operator, and therefore the absolute positions of the individual
components of the extended system, cannot be self-adjoint operators.

\vfill\break

\referenze{
\item{[1]} Luca Lusanna, {\sl Contemp. Math.}  {\bf 132}, 531 (1992).
\item{[2]} Luca Lusanna, {\it Dirac's Observables: from Particles to
Strings and Fields}, talk at the Int. Symposium on "Extended Objects
and Bound States", Karuizawa (Nihon Univ.) 1992.}

\vfill\break


\titolo{ Realization of $U_q(so(3))$ within $Diff({\bf R}_q^3)$}

\autori{\undertext{Gaetano~Fiore}}

As known, the $N$-dim real quantum euclidean space [1]
(${\bf R}_q^N$ in our notation) and its differential calculus [2] are
covariant w.r.t. the action of the quantum group of rotations $SO_q(N)$. Here
we restrict to the case $N=3$ and show that when $q\in {\bf R}$ it is possible
to realize $U_q(so(3))$ (which is
the dual Hopf Algebra of $Fun(SO_q(3))$) in terms of q-antisymmetric
differential operators on ${\bf R}_q^3$, i.e. in terms of the q-analog of the
angular momentum operators in ${\bf R}^3$. The explicit analogous
result for any
$N\ge 3$ will be given in [3] in detail.

The main result is the following. We define
$$
\cases{
L_m:=q[x^3\partial^1-x^1\partial^3+(q^{-{1\over 2}}-q^{1\over 2})x^2\partial^2]
B^{-1} \cr
L_+:=(q^{1\over 2}+q^{-{1\over 2}})^{1\over 2}q^{3\over 4}(x^2\partial^3-
qx^3\partial^2)B^{-1}  \cr
L_-:=(q^{1\over 2}+q^{-{1\over 2}})q^{1\over 4}(x^1\partial^2-qx^2\partial^1)
B^{-1}, \cr}                                                  \eqno (1)
$$
where for the ${\bf R}_q^3$ coordinates $x^i$ and the corresponding partial
derivatives $\partial^i$ we follow the conventions of [1],[2], and
$B:=1+q(q-1)x^i\partial_i$.
Then the generators $L_m,L_+,L_-$ make up a closed algebra with commutation
rules
$$
\cases{
q^{-{1\over 2}}L_mL_+-q^{1\over 2}L_+L_m=L_+    \cr
q^{1\over 2}L_mL_--q^{-{1\over 2}}L_-L_m=-L_-    \cr
[L_+,L_-]=(q^{1\over 2}+q^{-{1\over 2}})L_m[1+(q^{1\over 2}-q^{-{1\over
2}})L_m];
\cr}                                                         \eqno (2)
$$
moreover $(L_+)^*=L_-$, $L_m^*=L_m$, and the Casimir operator $L^2$ (the square
angular momentum) is the following quadratic expression in $L_+,L_-,L_m$
$$
L^2=L_m^2+(q^{1\over 2}+q^{-{1\over 2}})^{-1}
[q^{1\over 2}L_+L_-+q^{-{1\over 2}}L_-L_+].                    \eqno (3)
$$
In the classical limit $q=1$ we recover the usual $so(3)$ generators
$L_3,L_+,L_-$ satisfying the relations $[L_3,L_{\pm}]=\pm L_{\pm}$,
$[L_+,L_-]=2L_3$ after introducing suitable real coordinates.

The algebra generated by $L_m,L_+,L_-$ coincides with the Hopf algebra
$U_q(su(2))=U_q(so(3))$. In fact the transformation
$$
\cases{
L_{\pm}=(q^{1\over 2}+q^{-{1\over 2}})[2+(q^{1\over 2}-q^{-{1\over
2}})^2C]^{-1}
X_{\pm}q^{H\pm 1\over 4}   \cr
L_z=(q^{1\over 2}-q^{-{1\over 2}})^{-1}
\{-1+{
q^{H\over 2}(q^{1\over 2}+q^{-{1\over 2}})\over2+(q^{1\over 2}-
q^{-{1\over 2}})^2C}\}.\cr}                               \eqno (4)
$$
maps the set of generators $H,X_+,X_-$ of $U_q(su(2))$ satisfying the standard
[1] commutation relations (with $q\rightarrow q^{1\over 2}$)
into generators $L_m,L_+,L_-$ satisfying relations (2).

The unitary representation of integral spin $k$ of
$U_q(su(2))$ can be realized in terms of the differential operators
(1) acting on the $(2k+1)$-dimensional vector space $W_k$ made out of the
q-deformed
`` symmetric '' polynomials of degree $k$ in $x$. Then the scalar product
between
polynomials can be defined in terms of the integration on ${\bf R}_q^3$ defined
in [4],[5] by imposing Stoke's theorem. $W_k$ can be generated through
the iterated application of $L_+$ to the lowest weight eigenvector
$u_{k,-k}:=(x^1)^k$ of $L_m$. In fact the vectors
$u_{k,h}:=(L_+)^{k+h}
u_{k,-k}$,
$h=-k,-k+1,...,k$ ($L_+u_{k,k}=0$) make up a basis of $W_k$ consisting of
eigenvectors of $L^2,L_3$ with eigenvalues $l^2_k,\lambda_{k,h}$ respectively
given by
$$
l^2_k=[k]_q[k+1]_q\left({q^{1\over 2}+q^{-{1\over 2}}\over q^{k-{1\over 2}}
+q^{{1\over 2}-k}}\right)
$$
$$
\lambda_{k,-k}=-{1+q^{-1}\over q^k+q^{-k-1}}q^{1\over 2}[k]_q~~~~~~
\lambda_{k,h+1}=q\lambda_{k,h}+q^{1\over 2}.                  \eqno (5)
$$

\referenze{
\item{[1]} L.D. Faddeev, N.Y. Reshetikhin and L.A. Takhtajan, {\sl
Algebra and Analysis} {\bf 1} 178 (1989), translated
from the Russian in {\sl Leningrad Math. J.} {\bf 1}, 193 (1990).
\item{[2]} U. Carow-Watamura, M. Schlieker and S. Watamura, {\sl
Z. Phys. C; Part. Fields} {\bf 49}, 439 (1991).
\item{[3]} G. Fiore, in preparation.
\item{[4]} G. Fiore, {\it The $SO_q(N,{\bf R})$-Symmetric Harmonic
Oscillator
on the Quantum Euclidean Space ${\bf R}_q^N$ and its Hilbert Space
structure},
Sissa preprint 35/92/EP.
\item{[5]} A. Hebecker and W. Weich {\it Free Particle in q-deformed
Configuration Space}, LMU-TPW-1992-12.}

\vfill\break


\titolo{Quantum groups and lattice physics}
\autori{F.Bonechi, E.Celeghini, \undertext{R.~Giachetti},
E. Sorace and M.Tarlini}

{\bf 0.} A number of inhomogeneous quantum algebras of large physical interest
has recently been determined by extending the contraction procedure from
semisimple Lie algebras to their quantized version [1-3]. In order to preserve
the Hopf algebra structure, it may occurs that sometimes the quantum parameter
itself must undergo a rescaling, so that, after the contraction, the parameter
can acquire a physical dimension. It emerges then naturally a picture in which
the quantum algebra represents the symmetry of a dynamical system with a
fundamental length or time scale, $a$, and in which the coproduct $\Delta$
establishes the rules for combining the elementary excitations [4-6].

Here we shall review from this point of view the properties of {\it phonons},
which is the physical system associated with the two dimensional
pseudoeuclidean (or 1+1 -- Poincar\'e) group, $E_\ell(1,1)$, [5],
and then we describe the quantization of the linear modes of magnetic
chains, the {\it magnons}, by using a quantum analogue of the Galilei
group [6].
\smallskip
{\bf 1.} The quantum algebra $E_\ell(1,1)$ is generated by
$k^{\pm1}=e^{\pm iaP}$,
$P_0$, $J$, where $P$ represents the momentum, $P_0$ and  $J$ the energy and
boost respectively, satisfying\hfil\break
\centerline{$ kP_0k^{-1}=  P_0\,,\quad ~~~~~~~ kJk^{-1}=J+aP_0\,,\quad ~~~~~~~
JP_0-P_0J=(k-k^{-1})/(2a)\,,$}\hfil\break
$$\eqalign{{} & \Delta(k)   = k \otimes k\,,~~~~~~~~~~
  \Delta(P_0) = k^{-1/2}\otimes P_0 + P_0\otimes k^{1/2}\,,\cr
  {} & {}\phantom{XXXX}
\Delta(J)   = k^{-1/2}\otimes J  + J \otimes k^{1/2}\,.\cr}$$
The Casimir of $E_\ell(1,1)$ is
$\ C=P_0^2 - (2/a)^2 \sin^2(aP/2)\ $
and the differential realization
$P_0=(i/v)\ \partial_t\,, \quad   k=\exp(a
\partial_x)\,, \quad J=i(x/v)\partial_t-(vt/a)\sin(-ia\partial_x)\,,$
turns the eigenvalue equation $\,C\,z(x,t)=m^2v^2\,z(x,t)$
for the Casimir into the PDE\hfil\break
\centerline{$ \left(\partial_t^2 + (2v/a)^2 \sin^2(-ia\partial_x/2)
+ m^2v^4\right)\, z(x,t)=0\ , $}\hfil\break
which, for $m=0$, describes the phonons on a lattice with spacing $a$.
The single particle properties are well described by $E_\ell(1,1)$.
For instance the position operator
$X=(1/2)\ \bigl\{P_0^{-1},J\bigr\}_+$
has a time derivative
$v_g= \dot X = iv\,[P_0,X] = v\ \cos(aP/2)\,,$
reproducing the well known expression for the group velocity of the phonons.

The coproduct gives the two-phonon global variables. For
energy and boost we get\hfil\break
\centerline{$P_0 = e^{- i a P^{(1)}/2}\,P_0^{(2)} + P_0^{(1)}\,
e^{i aP^{(2)}/2}\,,~~~~~~~
J = e^{- i a P^{(1)}/2}\,J^{(2)} + J^{(1)}\,
e^{i a P^{(2)}/2}\,.$}\hfil\break
Moreover $k=k^{(1)}k^{(2)}\,$, from which
$P = P^{(1)} + P^{(2)} + 2\pi n/a$, whith $n$ chosen so that $P$
is kept in the fixed Brillouin zone: quantum symmetry implies Umklapp process.

Explicitly, for two possibly differently polarized phonons with velocity
parameters $v^{(1)},\ v^{(2)}$ and dispersion relations
$\ \Omega^{(r)} = v^{(r)}P_0^{(r)} = (2 v^{(r)}/a)\,\sin(aP^{(r)}/2)\,,$
$(r=1,2)$, the total energy $P_0$ reads
$\ P_0=(2/a)\sin(a(P^{(1)}+P^{(2)})/2)\ .$
Therefore, by the energy conservation, the physical process actually occurs
if there exists a velocity $v$ such that
$\Omega =\Omega^{(1)} + \Omega^{(2)}$,
where $\Omega=(2v/a)\sin(aP/2)$ is the dispersion relation of the composite
system.
\smallskip
{\bf 2.} A deformation $\Gamma_\ell(1)$ of the one-dimensional Galilei algebra
is obtained by the four generators $B$, $M$, $K$ and $T$ respresenting
the Galileian boost, the mass, the momentum and the energy which
satisfy
\hfil\break
\centerline{$[B,K]=iM\,,~~~~~~~[B,T]=(i/\ell)\,\sin(\ell K)\,,~~~~~~~
[M,\cdot]=[T,K]=0\,$}\hfil\break
$$ \eqalign{
   \Delta B=e^{-iaK}\otimes B
         +B\otimes e^{iaK}\,&,~~~~~~~~~~~~
   \Delta M=e^{-iaK}\otimes M
         +M\otimes e^{iaK}\,,\cr
   \Delta K = {\bf 1}\otimes K  +K\otimes{\bf 1}\,&,~~~~~~~~~~~~
\Delta T = {\bf 1}\otimes T  +T\otimes{\bf 1}\,,\cr}$$
The Casimir of $\Gamma_\ell(1)$ reads
$\ C=MT-(1/a^2)\;(1-\cos(aK))\ $
and this algebra admits the differential realization
$B=mx\,,~~~M =m\,,~~~P=-i\partial_x\,,~~~
T=(m a^2)^{-1}\Bigl(1-\cos(-ia\partial_x)\Bigr)+c/m\,,$
where $c$ is the constant value of the Casimir.

The physical system that can be studied by means of the quantum symmetry
$\Gamma_q(1)$ is a spin 1/2 system referred to as $XXZ$ model and known to
be integrable by the Bethe Ansatz method. Its Hamiltonian is
$ {\cal H}=2J\,\sum_{i=1}^N \Bigl(\,(1-\alpha)\,(S_i^x S_{i+1}^x +
               S_i^y S_{i+1}^y) + S_i^z S_{i+1}^z\,\Bigr)\ ,$
with $S_{N+1}^a=S_{1}^a$.
Firstly, from the Galileian position operator $X=B/M$, we find the
well known magnon velocity
$\dot{X}=i[T,X]=J\ell\,\sin(\ell K)$. Then,
following the standard method for analyzing such
models, the eigenvalue equation for ${\cal H}$, in
terms of the states $\psi=\sum_if_iS_i^+|0\rangle$ with a single spin deviate,
becomes an algebraic system in $f_i$
which can be embedded into the PDE for the continuous amplitude $f(x)$
\hfil\break
\centerline{$-4Js\Bigl(1-(1-\alpha)\cos(-ia\partial_x)\Bigr)f(x)=
  (\epsilon-\epsilon_0)f(x)\ .$}\hfil\break
The operator on the left hand side coincides with the differential realization
of $T$, by identifying
$(m\ell^2)^{-1}=-2J(1-\alpha)$ and  $c/m=-2J\alpha\,.$

Let us now discuss the two magnon states $\psi=\sum_{ij}f_{ij}S_i^+S_j^+
|0\rangle$, where $f_{ij}=f_{ji}$, $i\not= j$, while $f_{ii}$ are
physically meaningless and have no part in the theory. The algebraic system for
the coefficients $f_{ij}$ is easily found and can again be analyzed by
embedding it into a PDE for continuous amplitudes $f(x_1,x_2)$ and using
the $\Gamma_q(1)$ symmetry. From $\Delta T$ we find the total energy
$
T_{12}
= (M_1 \ell^2)^{-1} \Bigl(1-\cos(\ell K_1)\Bigr) +
(M_2 \ell^2)^{-1}\Bigl(1-\cos(\ell K_2)\Bigr)  + (c_1/M_1) + (c_2/M_2)\,.$
For $M_1=M_2=M$, using the previous identifications and the differential
realization $K_1=-i\partial_{x_1}\,,~K_2=-i\partial_{x_2}\,,$ the eigenvalue
equation $T_{12}f(x_1,x_2)=(\epsilon-\epsilon_0)f(x_1,x_2)$ for the two magnon
amplitude $f(x_1,x_2)$ is equivalent to the free system of the Bethe ansatz.
The two magnon bound states are obtained by requiring that the energy has a
homogeneous dependence of degree $-1$ upon the total mass, exactly as in the
single magnon cases. As a result we obtain a total mass
$M_{12}=2M/(1-\alpha)$ and the energy of bound states is
$T_{12}=-2J\Bigl(1-(1-\alpha)^2\,\cos^2(\ell K/2)\Bigr)\,.$

The procedure can be extended to
any number of magnons by using the coproduct and its associativity:
we find two recurrence relations for the $n$-magnon
mass and energy that can be solved, yielding\hfil\break
\centerline{$M_{12\dots k} =-\Bigl(2J(1-\alpha)\ell^2\Bigr)^{-1}\,
{\cal U}_{k-1}(1/(1-\alpha))\ ,\quad k=2,\dots n\ ,$}\hfil\break
\smallskip\noindent
\centerline{$T_{12\dots n}=
{{\strut\displaystyle {-2J(1-\alpha)}\over\displaystyle
{{\cal U}_{n-1}(1/(1-\alpha))}}}\,\Bigl(
   {\cal T}_n(1/(1-\alpha)) - \cos(\ell K_{12\dots
n})\Bigr)\,,$}\hfil\break
so that the bound state energy of the $n$ magnon bound states has a closed
form in terms of the Tchebischeff polynomials ${\cal U}_k$ and ${\cal T}_k$.

\referenze{
\item{[1]} E. Celeghini, R. Giachetti, E. Sorace and M. Tarlini, {\sl J. Math.
      Phys.} {\bf 31}, 2548 (1990); {\sl J. Math. Phys.}
      {\bf 32}, 1155 (1991);
      {\sl J. Math. Phys.} {\bf 32}, 1159 (1991);
      {\it Contractions of quantum groups}, in {\sl Quantum Groups},
      Lecture Notes in Mathematics n. 1510, 221, (Springer-Verlag, 1992).
\item{[2]} E. Celeghini, R. Giachetti, A. Reyman, E. Sorace and M. Tarlini,
      {\sl Lett. Math. Phys.} {\bf 23}, 45 (1991).
\item{[3]} E. Celeghini, R. Giachetti, P. Kulish, E. Sorace and M. Tarlini,
      {\sl J. Phys.} {\bf A24}, 5675 (1991).
\item{[4]} E. Celeghini, R. Giachetti, E. Sorace and M. Tarlini,
      {\sl Phys. Lett.} {\bf B280}, 180 (1992).
\item{[5]} F. Bonechi, E. Celeghini, R. Giachetti, E. Sorace and M. Tarlini,
      {\sl Phys. Rev. Lett.} {\bf 68}, 3718 (1992).
\item{[6]} F. Bonechi, E. Celeghini, R. Giachetti, E. Sorace and M. Tarlini,
      {J. Phys.} {\bf A25}, L939 (1992); {\sl Phys. Rev.} {\bf B46},
      5727 (1992).}

\vfill\break


\titolo{Toda field theories and quantum groups}

\autori{\undertext{Loriano~Bonora}}

The simplest Toda theory in 1+1 dimensions is the Liouville theory.
The
Liouville equation
$$ \partial_{x_+ }\partial _{x_-}\varphi= e^{ 2\varphi},\quad
\quad \quad \quad x_\pm=x\pm t,\quad\quad x\in {\bf  S^1}, \,t\in {\bf R}
$$
is conformal invariant. It can be solved through the following
procedure.
Write the associated
Drinfeld--Sokolov linear systems
$$
\partial_{x_+}Q(x_+)= \Big( p(x_+)H-E_+\Big) Q(x_+),\quad\quad
\partial_{x_-}\bar Q(x_-)= -\bar Q(x_-) \Big(\bar p(x_-)H-E_-\Big)
$$
where $H,E_+,E_-$ are the generators of the Lie algebra $sl_2$.
For any solutions $Q_+$
and $Q_-$ of these systems and any highest weight vector $|
\Lambda >$ of $sl_2$
corresponding to the weight $\lambda$
$$
e^{\lambda (\varphi(x_+,x_-) H)}= <\Lambda|Q(x_+)M\bar Q(x_-)|\Lambda>
$$
is a solution of the Liouville equation for any constant matrix $M$.
The matrix $M$ is
chosen in such a way as to guarantee periodicity and locality of the solutions.
Vice versa one can prove that to any local periodic solutions of the
 Lioville  equation
there corresponds a couple of free periodic chiral boson fields
 $p$ and $\bar p$.
This allows us to describe the classical phase space of the Liouville
theory by means
of free bosonic oscillators [1].

This analysis can be extended to any Toda field
theory based on a finite dimensional Lie algebra [2], to Toda field
theories based
on affine algebras [3] and to Toda field theories defined on Riemann
surfaces [4].

The particularly simple parametrization of the classical phase space
(free bosonic
oscillators) makes the canonical quantization procedure very effective. An
intermediate quantum formula is, for example, the exchange algebra
$$
\psi(x) \psi(y) = \psi(y) \psi(x) R^\pm_{12}(p_0), \quad\quad  +(-) \quad
{\rm if}\quad x >(<) y
$$
where $\psi= <\Lambda| g\rho$, and $g, \rho$ are suitable matrices that
guarantee periodicity and locality. $R_{12}$ is the quantum $R$ matrix
in the Block wave basis [5,6].

Quantization of the Liouville theory eventually leads to calculating the
correlation functions in particular of the conformal minimal models.
Quantization of the other Toda theories based on finite dimensional
Lie algebras
leads to W minimal models, and finally quantization of Toda theories based
on affine Lie algebras should lead to a thorough understanding of theories
like the sine--Gordon or sinh--Gordon models and their generalizations.

\referenze{
\item{[1]} E. Aldrovandi, L. Bonora, V. Bonservizi, R. Paunov,
{\it Free field representation
    of Toda field theories} preprint SISSA 210/92/EP.
\item{[2]} O. Babelon, L. Bonora, F. Toppan, {\sl Comm. Math. Phys.} {\bf
140}, 93(1991).
\item{[3]} O. Babelon, L. Bonora, {\sl Phys. Lett.} {\bf B244}, 220
(1990);
{\bf B267}, 71 (1991).
\item{[4]} E. Aldrovandi, L. Bonora, {\it Liouville and Toda field
theories
on Riemann surfaces}
  preprint SISSA 27/93/EP.
\item{[5]} O. Babelon, L. Bonora, {\sl Phys. Lett.} {\bf B253}, 365 (1991).
\item{[6]} L. Bonora, V. Bonservizi, {\it Quantum $sl_n$ Toda field
theories}, preprint SISSA 110/92/EP, to appear in Nucl. Phys. B.}

\vfill\break


\titolo {``Embedding" of q-algebras}

\autori{\undertext{Antonio~Sciarrino}}

The underlying idea in some
applications of q-algebras is to use a q-deformed algebra
instead of a Lie algebra to realize a {\it dynamical symmetry}.
The dynamical symmetry in many physical models is displayed through embedding
chains of algebras of the type
$$
G_0 \supset G_1 \supset \ldots \supset SO(3) \supset SO(2)
$$
An essential step to carry forward the program of application of
q-algebras as {\it dynamical symmetry} is to dispose on a formalism which
allows to build up chains analogous to eq.(1)
replacing the Lie algebras by the deformed ones.
The existence of 3-dim principal q-subalgebra for $Gl_q(3)$,
has been shown [1] in the symmetric basis, but the the coproduct of $Gl_q(3)$
{\bf does not induce} a coproduct in the 3-dim principal subalgebra.
In ref.[2] we tried to solve the problem the other way around:
to define $SO_q(3)$ and to build up a deformed structure of the type
$Gl_q(3)$. Indeed a ``deformed $~Gl(3)$" can be obtained but it is
not clear how to impose on it the Hopf structure.
The origin of the difficulties is on the Chevalley-Cartan basis which
is not suitable to discuss embedding of subalgebras except the trivial ones.
We present here an alternative deformation scheme [3].
Let us immediately emphasize that really the word ``embedding" is
used in some loose sense: the algebra $G_q$ deformed according the following
deformation scheme is {\bf not} the same as the $G_q$ defined in the Chevalley
basis. Let us sketch what the underlying idea is. Consider a semisimple Lie
algebra $G$ and a not regular maximal subalgebra $ L \subset G $:
$ ad_G \rightarrow ad_L \oplus R_L $ where $R_L$ is a representation of $L$.
Let $\{E_i^{\pm}\}$ be the generators of $L$ in the Chevalley basis and
$\{X_k^{\pm}\}$ some elements of $R_L$ with suitable properties.
Then we define a deformation scheme in which the Cartan subalgebra of $~G~$
, which is partly in the Cartan subalgebra of $~L~$ and partly in $~R_L~$,
is left invariant; the set of $\{E_i^{\pm}\}$ is deformed in the
standard way.
In the simplest case where the rank of $L$ is one unit less
the rank of $G$ there is one element $K_0$ in $R_L$ which commutes
with the Cartan subalgebra of $L$ such that (for fixed $ j $, $ i \neq j $ ):
$$
{}~[K_0,\, E_j^{\pm}] = \pm X_j^{\pm} ~~~~~ [K_0,\, E_i^+]\,] = 0
{}~~~~~ [X_j^+,\, X_j^-]  = [H_j]_{q_j}
$$
Then we impose the Hopf structure on $K_0$ as an element of the Cartan
subalgebra. This scheme defines $ G_q \supset L_q $.

\referenze{
\item{[1]} Van der Jeugt, {\sl J. Phys.} {\bf A25}, L213 (1992).
\item{[2]} A. Sciarrino, {\it Deformed U(Gl(3))} from $SO_q(3)$, Proc.
              Symmetries in Science VII: Spectrum Generating Algebras and
              Dynamics in Physics, Ed. B. Gruber, Plenum Prees Pub., to be
              published.
\item{[3]} A. Sciarrino, {\it Deformation of Lie algebras in a
non-Chevalley basis and embedding of q-algebras, DSF-Preprint},
preprint Universit\`a di Napoli 1993.}

\vfill\break


\titolo{Anyons and Quantum Groups}

\autori{Alberto Lerda and \undertext{Stefano~Sciuto}}

Anyonic oscillators with fractional statistics are built
on a two-dimensional square lattice by means of a generalized
Jordan-Wigner construction, and their deformed commutation
relations are thoroughly discussed. Such anyonic oscillators,
which are non-local objects that must not be confused with
$q$-oscillators, are then combined \`a la Schwinger to
construct the generators of the quantum group $SU(2)_q$
with $q=\exp({\rm i}\pi\nu)$, where $\nu$ is the anyonic
statistical parameter [1].

The construction can be generalized to $SU(n)_q$ [2] and more generally
to the deformations of the Lie algebras $A_n$, $B_n$ and $C_n$, which can be
built by a fermionic Schwinger construction [3].

\referenze{
\item{[1]} A. Lerda and S. Sciuto, {\it  Anyons and Quantum Groups},
Torino and Stony Brook preprint DFTT 73/92, ITP-SB-92-73, December 1992.
\item{[2]} R. Caracciolo and M.A. R-Monteiro, {\it Anyonic
Realization of $SU_q(N)$ Quantum Algebra}, Torino preprint DFTT 5/93,
February 1993.
\item{[3]} M. Frau and M.A. R-Monteiro, in preparation.}

\vfill\break


\titolo{ From classical to quantum: the problem of universality    }

\autori{\undertext{Piero~Truini}}

A basic question for a theory
is that of {\it universality}: one should be able to build within a certain
class of objects a general structure which incorporates every model of that
class as a particular case. We have addressed this question
within the theory of Quantum Groups, in particular, for
deformations of the universal enveloping algebras of reductive Lie
algebras (a Lie algebra is {\it reductive\/} if it is the direct
sum of a semisimple Lie algebra and an abelian one,  as in the case of
$GL(n)$).

We have solved the question, [1,2] starting from very simple concepts and
an Ansatz. The main results of our work can be summarized as follows.
Given an algebra or a Hopf algebra $A$ over $\bf C$ we define
a deformation of it as an algebra or a Hopf algebra
over $R={\bf C}[\! [u_1,u_2,\dots ,u_m]\! ]$, the ring of formal power series
in the deformation parameters $u_1, u_2, \dots ,u_m$,  which is
torsion
free as
an $R$-module, that reduces to $A$
when $u_1=u_2=\dots =u_m=0$. The number of deformation
parameters is arbitrary to start with.

The first important step towards our goal is the construction of what
we call {\it the extended enveloping algebra}, which is an
enlargement of the classical enveloping algebra (but where the algebra
structure is deformed) so as to include power series in the Cartan
generators. This leads to a result that we regard as fundamental in the
characterization of quantum groups.
The deformations of universal enveloping algebras involve some infinite
series in the generators of the Cartan subalgebras, but
we find the characterization of such series, given in the
literature, quite unsatisfactory and sometimes not fully justified.
We show that these series are strongly constrained. The coalgebra
structure forces them to satisfy
harmonic, constant coefficient differential equations; the algebra
structure implies invariance under translations. Both constraints
together leave as possibilities only certain combinations of
polynomials and exponentials.

Having established this we prove, under the Ansatz that the two Borel
subalgebras deform as Hopf algebras, that all deformations of a reductive
Lie algebra can be obtained
as specializations of a universal multiparameter deformation. The
number of parameters in the universal deformation is ${1\over
2}[N(N-1)-C(C-1)]+M$, where $N$ is the rank of the reductive algebra, and
$C=N-N_1$ is the dimension of its center, $N_1$ being the rank of the
semisimple part, while $M$ is the number of simple components.

We also prove that the algebra structure can always be reduced (on the simple
components) to that of the standard one-parameter quantization that
can be
found
in the literature. The extra parameters then appear only
in the comultiplication. If we restrict our result to the case of a simple
algebra, we find precisely the twisted quasi triangular Hopf algebras,
the twisting being a particular case of {\it gauge} transformation in the
category of quasi triangular quasi-Hopf algebras.

The example of $A_1\oplus A_1$, that we had treated in [3], having in mind
there non-standard deformations of the Lorentz group ($A_1\oplus A_1$ is the
complexification of its Lie algebra), serves
as an illustration of the nature of the extra parameters.
This example suggests that part of them
is closely related to quantizations of a semisimple algebra in which
the simple components remain classical throughout the deformation.

In a recent paper, [4], we found the dual structure of the
universal deformations found in [1,2], in the case
of semisimple A-type Lie algebras. In particular we can define, through a
pairing, a natural action of the universal deformation of the Lorentz algebra
on the deformed Minkowski space that suggests a new way of deforming the
Poincar\'e Lie algebra.

\referenze{
\item{[1]} P. Truini and V.S. Varadarajan,
          {\sl Lett. Math. Phys.}, {\bf 26}, 53 (1992).
\item{[2]} P. Truini and V.S. Varadarajan, GEF-Th-14/1992, to
          appear in {\sl Rev. Math. Phys.}.
\item{[3]} P. Truini and V.S. Varadarajan,
          {\sl Lett. Math. Phys.}, {\bf 21}, 287 (1991).
\item{[4]} E. De Vito and P. Truini GEF-Th-3/1993, submitted for
          publication.}

\vfill\break


\titolo{Anyons in planar ferromagnets}

\autori{Luigi Martina, \undertext{Oktay~Pashaev} and  Giulio Soliani }

As is 	well known, the Gauss law
constraint in  the  Chern-Simons ($CS$) theory can be related
to   the  comultiplication  of
a quantum group [1]. This  emerges  as  a  hidden  symmetry
of  the quantized $CS$
theory [2].  Correspondingly, the  soliton (vortex)  excitations  at
the quantum level satisfy an arbitrary statistics and can be
regarded as  anyons,  identified with Laughlin's
quasiparticles in the Quantum Hall Effect  [3]. \par We show that,
at least at the classical  level,  a planar  continuum Heisenberg
ferromagnet, given
 by the Landau-Lifshitz equation $$ {\bf
S}_{t} = {\bf S \times }\, (\partial ^{2}_{1}+\partial
^{2}_{2})\,{\bf S},  \eqno (1) $$ \noindent where ${\bf S} = {\bf
S}(x_{1}, x_{2},t), {\bf S}^{\,2}= 1$, can be described in terms of
the  $CS$ theory. \par By resorting to  the tangent space
representation [4],  in  the  {\it static case}  Eq. (1)
can be reduced to the (anti) self-dual $CS$ model \par $$
\eqalign{D_{\mp }\psi _{\pm }\;\;\;& =\; 0 ,  \cr
[D_{1},D_{2}]&=4i\kappa ^{2}(\mid \psi _{+}\mid ^{2}- \mid \psi
_{-}\mid ^{2})\;.}\eqno (2) $$ \par   If, for instance,   $\psi
_{- }$  is  vanishing,  then
$\mid\psi_{+} \mid^2$  satisfies the Liouville equation
$$\left({{{\partial }_{1}}^{2}+{{\partial }_{2}}^{2}}\right)\ell
n{\left|{\psi_{+} }\right|}^{2}=-8{\kappa }^{2}{\left|{\psi_{+}
}\right|}^{2}.\eqno (3)$$ Thus,  the magnetic vortices of model
(1) correspond to the $CS$ and Liouville solitons, while the
topological charge of  the former corresponds  to the electric
charge of the latter. \par  If both  the functions $\psi _{\pm }$
are nonvanishing [5],  then we can introduce a holomorphic
function $U = \overline{\psi }_{+} \,\psi _{-}$  and  rewrite  Eq.
(2) as the conformally invariant Sinh - Gordon equation\par
$$\eqalign{ (\partial ^{2}_{1}+ \partial ^{2}_{2})\ln \mid \psi
_{+ }\mid ^{2} &= -8\kappa ^{2}(\mid \psi _{+ }\mid ^{2}-
\mid U\mid ^{2}/\mid \psi _{+ }\mid ^{2}),\cr  (\partial
^{2}_{1}+ \partial ^{2}_{2})\ln \mid U\mid ^{2}\; &= \;0\qquad ,}
\eqno(4) $$ which can be extended to the affine Liouville model [6]. \par In
the   {\it non-static case},  following the same procedure,
we find a $CS$   gauged
Nonlinear Schr\"odinger Equation for the  charged matter  fields
$\psi_{\pm}$ coupled to
a "statistical gauge field" $A_{\mu}$ [5], namely $$\eqalign{
iD_{0}\psi _{\pm } + (D^{2}_{1}&+ D^{2}_{2})\psi _{\pm } + 8\kappa
^{2}\mid \psi _{\pm }\mid ^{2}\psi _{\pm } = 0,\cr \partial
_{1}A_{2} - \partial _{2}A_{1}& = -8\kappa ^{2}(\mid \psi
_{+}\mid ^{2}- \mid \psi _{-}\mid ^{2}), \cr \partial
_{0}A_{i}- \partial _{i}A_{0}&= 8\kappa ^{2}i\epsilon
_{ij}{(\overline{\psi }_{+} D_{j} \psi_{+} - \overline{D_{j} \psi_{+}}\,\psi
_{+}) - (\overline{\psi }_{-} D_{j}\psi _{-} -  \overline{D_{j} \psi_{-}}\,\psi
_{-})}}$$ and   connected by the
relation    $$D_{+}\psi _{-} = D_{-}\psi _{+}, $$ where $
D_{\mu } = \partial _{\mu } - i/2 A_{\mu },\;  D_{\pm }\equiv
D_{1}\pm  iD_{2},\; A_{0}= V_{0} - 8\kappa
^{2}(\mid \psi _{+}\mid ^{2}+ \mid \psi _{-}\mid ^{2}), \;
A_{i}\equiv   V_{i} \;(i=1,2),$  and $V_{\mu }$ is  the $U(1)$
gauge field associated with the tangent space.

This work was supported by MURST
of Italy.

\referenze{
\item{[1]}  E. Witten, {\sl Comm. Math. Phys}, {\bf 121}, 351 (1989);
           E. Witten, {\sl Nucl. Phys.} {\bf B 330}, 285 (1990).
\item{[2]} J. Fr\"olich, in Carg\'ese Lectures 1987.
\item{[3]}  R.E. Range and S.M.
Girvin, {\it  The quantum Hall Effect}  (Springer-Verlag, New York,
1990).
\item{[4]} L. Martina, O.K. Pashaev and G. Soliani, {\it Self-Dual Chern-Simons
Solitons in Nonlinear $\sigma$-Models
}, Preprint University Lecce, January 1993.
\item{[5]}  L. Martina, O.K. Pashaev and G. Soliani, {\it Chern-Simons Gauge
Theory of Planar  Ferromagnets},  Preprint University Lecce, in preparation.
\item{[6]} O. Babelon and L. Bonora, {\sl Phys. Lett.},
{\bf B 244}, 220 (1990). }

\vfill\break


\titolo{Positive energy unitary representations of the conformal q-algebra}

\autori{\undertext{Ludwik~Dabrowski}}

We report the work [1] on
positive-energy unitary irreducible representations of a q-deformation
${\cal U}_q(su(2,2))$
of the conformal algebra, which has the $q$-deformed
Lorentz algebra ${\cal U}_q(so(3,1))$ as a Hopf subalgebra [2].
For generic $q$, the positive energy UIRs of ${\cal U}_q(su(2,2))$
are deformations of the respective representations of $su(2,2)$
labelled by $[j_1,j_2,d]$, where $2j_1, 2j_2$ are non-negative
integers fixing finite dimensional irreducible representations of
the Lorentz subalgebra $so(3,1)$, and ~$d$~ is the conformal
dimension [3].
We extend for $|q| = 1$, the hermitian conjugation used in [3] to an
anti-linear anti-involution $\omega $ in a $*$-Hopf algebra
${\cal U}_q(sl(4,C))$,
which is a complexification of ${\cal U}_q(su(2,2))$.
With the help of $\omega $ we introduce the scalar product of the
Poincar\'e-Birkhof-Witt basis in the Verma module $V_q$.
By using some new results in the theory of Verma modules [4],
we find the singular (null) vectors and after quotienting them and their
descendents in $V_q$, we obtain irreducible
positive-energy unitary representations.
When $q$ is $N$-th root of unity, all
these unitary representations become finite-dimensional.
We discuss in some detail the massless representations,
which are also irreducible representations of the $q$-deformed
Poincar\'e subalgebra. Generically, their dimensions
are smaller than the corresponding finite-dimensional non-unitary
representation of $su(2,2)$, except when $N = 2|h|+1$, where
$h$ is the helicity (this includes the fundamental representations).
As examples we give explicitly a $4\times 4$ representation for
$N = 2\ (q = -1), j_1 = 1/2, j_2 = 0$,
and list the orthonormal basis for the cases
$N = 3\ (q = e^{2\pi i/3}), j_2 = 0, j_1 = 1/2$ and $j_1 = 1$.

\referenze{
\item{[1]}
L. Dabrowski, V.K. Dobrev, R. Floreanini and V. Husain,
{\sl Phys. Lett. B}, in print
\item{[2]} V.K. Dobrev, {\it Canonical $q$-deformations of
noncompact Lie (super-) algebras},\hfill\break G\"ottingen University
preprint, (July 1991), to appear in {\sl J. Phys. A}; ~cf. also:
{\it $q$-Deformations of noncompact  Lie (super-) algebras: the
examples of $q$-deformed Lorentz, Weyl, Poincar\'e and (super-)
conformal algebras}, ICTP preprint, IC/92/13 (1992), to appear in
the Proceedings of the Quantum Groups Workshop of the Wigner
Symposium (Goslar, July 1991).
\item{[3]} G. Mack, {\sl Commun. Math. Phys.} {\bf 55}, 1 (1977).
\item{[4]} V.K. Dobrev, ICTP Trieste internal report IC/89/142
(June 1989) and in {\it Proceedings of the International Group
Theory Conference} (St. Andrews, 1989), Eds. C.M. Campbell and
E.F. Robertson, Vol. 1, London Math. Soc. Lect. Note Ser. 159
(1991) p. 87.}

\vfill\break


\def\endo#1{\mathop{End(\complessi^#1\otimes\complessi^#1)}}
\def\complessi{\mathinner{\bf C}}

\titolo{Construction of quantum groups and the Y.B.E. with
spectral parameter}

\autori{\undertext{P.~Cotta~Ramusino}, L. Lambe and M. Rinaldi}

Faddeev, Reshetikhin and Takhtadzhyan showed, some time ago
[1]
how to construct a bialgebra $A_R$ out of a given matrix $R\in
\endo{n}$ satisfying the Yang-Baxter equation. This bialgebra
may have an antipode, i.e. may be  Hopf algebras, when some
extra conditions are satisfied.
\par
All known Quantum Groups arise as Hopf algebras which are
duals to the above bialgebras.
In this framework, the Universal R-Matrix
can be thought as a bilinear form
over $A_R$ [2].
\par
In our paper we analyze some aspects of the construction
of [1] and discuss some applications
to the case of Yang-Baxter
equations with spectral parameter.
\par
We first start with a
matrix $R \in \endo{n}$, which is only supposed to verify the
Yang-Baxter equation (without spectral parameter)
and then
we construct canonically a pair of bilinear
forms on $A_R$ satisfying the quasi-commutativity and quasi-triangularity
properties, as well as the Yang-Baxter equation.
\par
These bilinear forms (Generalized Universal R-Matrices) allow the
construction of other solutions of the Yang-Baxter equations which can be
seen as ``tensor products" of the given solution.
\par
When $R$,
$R^{t_2}$ and $(R^{-1})^{t_2}$ are all
invertible then
we show explicitly
that the dual
of $A_R$ is a ribbon Hopf-Algebras
(see [3,4]).
Above, given $R = \sum_i \alpha_i \otimes \beta_i,$ we have set
$R^{t_2}
\equiv \sum_i \alpha_i \otimes \beta^T_i.$

\par
The construction of Generalized Universal R-Matrices,
can be easily adapted to the case of Yang-baxter equations with
spectral parameter. Moreover the spectral parameter can take values in
a generic non commutative group. \par
If we are given a solution of such of an equation, then we can
construct in a canonical way a bialgebra, very similar to
the bialgebra $A_R$ considered in the constant case and a
pair of bilinear forms over such bialgebra. These
bilinear forms have all the corresponding properties of
the constant case (quasi-commutativity, quasi-triangularity)
and they satisfy the Yang-Baxter equation {\it without} the spectral
parameter. \par
This allows us to construct  tensor products of solutions,
exactly as in the constant case. Moreover we
discuss the search for solutions of the
Yang-Baxter equation with spectral parameter, starting from the
solutions of the corresponding constant equation.

\referenze{
\item{[1]} N. Reshetikhin, L. Takhtadzhyan, L. Faddeev, {\sl
Leningrad Math. Journ.}, {\bf 1}, 193 (1990).
\item{[2]} S. Majid, {\sl Intern. Journ. Mod. Phys.}, {\bf A5}, 1 (1990).
\item{[3]} V. Turaev,{\sl Invent. Math.} {\bf 92}, 527 (1988).
\item{[4]} N. Reshetikhin, {\sl Leningrad Math. Journ.}, {\bf 1},
491 (1990).}

\vfill\break


\titolo{Topological field theory and invariants of three-manifolds}

\autori{\undertext{Enore~Guadagnini}}

The quantum Chern-Simons model [1,2] is a topological gauge
field theory which is defined
in a generic three-manifold $\cal M$ which is orientable.  The gauge invariant
observables of this system are the vacuum expectation values of the Wilson line
operators associated with framed and oriented links in $\cal M$;
each link component has
a colour which is given by the labels of the inequivalent irreducible
 representations of
the gauge group.

Quite remarkably, for compact simple Lie groups, the Chern-Simons theory
is soluble [3] in any closed and connected three-manifold.
This means that, for any given
link in $\cal M$, one can easily compute the exact expression of the associated
expectation value of the Wilson line operator.
These expectation values represent
invariants of ambient isotopy for framed links in $\cal M$.

In addition to the expectation values of the Wilson line operators,
one can also
compute the value ${\cal I}({\cal M})$ of the improved partition
function [3] of the
theory in the three-manifold $\cal M$. In order to give a precise
definition of ${\cal
I}({\cal M})$, one must connect topologically inequivalent manifolds. This
problem cannot be solved by means of the standard methods of field
theory.
For this
reason, in ref.[3] the operator surgery  method has been introduced.
The basic idea is
to find an appropriate generalization of the familiar Casimir effect.
Any three-manifold
$\cal M$ is cobordant with the sphere $S^3$; this means that $\cal M$
can be obtained by
means of Den surgery on $S^3$. The operator surgery method consists
in finding the
field theory rules which correspond to the surgery instructions.
The crucial property,
which permits us to solve this problem, is the fact that a generic
surgery can be obtained
by combining twist homeomorphisms of solid tori.

The case in which the gauge group is $G=SU(2)$ has been studied in
detail in ref.[3];
the operator surgery method can also be used for more complicated
groups.
So, the
non-Abelian Chern-Simons field theory (with a compact gauge group)
has been solved in any
three-manifold $\cal M$.  As a result, one can define and compute a
set of topological
invariants of three-manifolds. These invariants are equivalent to
those defined by
Reshetikhin and Turaev [4] by means of quantum groups. This
equivalence is not a
coincidence because any function, which is invariant under the
diagonal
action of the
standard quantum groups, really represents an invariant function of
ordinary Lie groups.

Given a set of topological invariants in three dimensions, one can
naturally define a
set of associated invariants in two dimensions. Thus, by means of the
invariants of the
Chern-Simons field theory, one can define [5] a set of topological
invariants for the
punctured Riemann surfaces of arbitrary genus. These new invariants
are integer numbers
and represent the dimensions of the physical state spaces associated
with the punctured
Riemann surfaces.

\referenze{
\item {[1]} E. Witten, {\sl Commun. Math. Phys.} {\bf 121}, 351 (1989).
\item {[2]} E. Guadagnini, {\sl Int. Journ. Mod. Phys.} {\bf A7},
877 (1992);
E. Guadagnini, {\sl Phys. Lett.} {\bf B260}, 353 (1991).
\item {[3]} E. Guadagnini,  {\sl Nucl. Phys.} {\bf B375}, 381 (1992);
E. Guadagnini and S. Panicucci,  {\sl Nucl. Phys.} {\bf B388}, 159 (1992).
\item {[4]} N.Y. Reshetikhin and V.G. Turaev, {\sl Commun. Math.
Phys.} {\bf 127}, 1 (1990);N. Reshetikhin and V.G. Turaev, {\sl
Invent. Math.} {\bf 103}, 547 (1991).
\item {[5]} E. Guadagnini, {\it Topological Invariants in Two and
Three Dimensions}, Preprint University of Pisa, IFUP-TH.11/93.}

\vfill\break


\titolo { The quantum Dirac equation associated to the $\kappa$--Poincar\'e}

\autori{Anatol Nowicki, \undertext{Emanuele~Sorace} and Marco Tarlini}

In these last years the method of "$q$--contracting" known
semisimple quantum groups to nonsemisimple
ones [1] has been successfully applied to get the deformed
counterpart of the most physically relevant kinematical algebras [2,3].
The construction of 4--dimensional quantum Poincar\'e (with 10
parameters and an hermitian involutive coproduct) [3,4] which has been
obtained by  $q$--contracting the $q$--antiDeSitter $SO_q(3,2)$ Hopf algebra
rewritten in terms of $q$--commutators of "physical generators".
Since only one direction can be deformed it is obvious to choose the
time axis so that the quantum relativistic mass square
operator becomes $C_1 = (2\kappa \sinh(P_0/(2\kappa))^2-P_iP_i$
(as proposed in [2]) and the square quantum Pauli--Lubanski operator
is [4,5]
$ C_2=(\cosh(P_0/\kappa)-
{P_iP_i}/(4\kappa^{2}))\; W_0^{2}- W_iW_i,\ $
with
$W_0=P_i M_i$ and
$ W_i=\kappa \sinh(P_0/\kappa)M_i + \epsilon_{ijk}P_jL_k $.

A $q$--deformed Dirac equation
was then proposed in [5,6] as the square root of $C_1$.
Our criterion [7] is  that the $\kappa$--deformed Dirac equation must be
invariant under the spinorial representations of the $\kappa$--Poincar\'e
and our method makes use of the contraction applied to representations.
So we $q$--contract the sum of two $q$--antiDeSitter representations
one of them 4--dimensional, rephrasing the procedure that in the
classical case gives rise to the Dirac equation (see [8]).
In this way we derive the corresponding representation of the
$\kappa$--Poincar\'e generators in which the
orbital spinless operators are q-composed with the $ \gamma$
representation of the Lorentz algebra (which is a zero--momentum
representation also of the $\kappa$--Poincar\'e Hopf algebra).
For these representations the relationship between the two Casimirs is
$ C_2 = -(3/4) C_1(1 + C_1/(4\kappa^2))$.

The quantum Dirac operator is then found by imposing the invariance
under the global $\kappa$--spinorial representations:
$$D= - \exp(-{P_0\over2\kappa})\,
\gamma_i\, P_i + \gamma_4\; \kappa\sinh({P_0\over\kappa})
 -\gamma_4\, {1\over2\kappa}\, P_i P_i\ .$$
It is noteworthy that it differs from [5,6] as
$ D^{2}= C_1(1 + C_1/(4\kappa^2))= -(4/3)\, C_2$ and that $D$ gives
rise to a 4--spinorial wave equation with finite--difference operators
in the time with a delay $i/(2\kappa)$, the classical limit $\kappa
\rightarrow \infty$ gives the standard Dirac operator.
The phenomenological implications concerning the coupling to electromagnetic
field are now under investigations [9,10].

\referenze{
\item{[1]} E. Celeghini, R. Giachetti, E. Sorace and M. Tarlini,
      {\sl J. Math. Phys.} {\bf 32}, 1155 (1991).
\item{[2]} E. Celeghini, R. Giachetti, E. Sorace and M. Tarlini,
      {\sl J. Math. Phys.} {\bf 32}, 1159 (1991);
      {\it Contractions of quantum groups}, in
      ``Quantum Groups'',Lecture Notes in Mathematics n.1510, 221,
      (Springer-Verlag, 1992).
\item{[3]} J. Lukierski, A. Nowicki, H. Ruegg and V.N. Tolstoy, {\sl
 Phys. Lett.} {\bf B 264}, 331 (1991).
\item{[4]}J. Lukierski, A. Nowicki, H. Ruegg, {\sl Phys.Lett.} {\bf B293},
344 (1992).
\item{[5]} S. Giller, J. Kunz, P. Kosinski, M. Majewski and P. Maslanka,
      {\sl Phys.Lett.} {\bf B286}, 57 (1992).
\item{[6]} S. Giller, J. Kunz, P. Kosinski, M. Majewski and P. Maslanka,
      {\it On q-covariant wave functions}, Lodz University Preprint,
      August 1992.
\item{[7]} A. Nowicki, E. Sorace and M. Tarlini, {\it The quantum deformed
      Dirac equation from the $\kappa$--Poincar\' e algebra}. Firenze Preprint,
      December 1992, {\sl Phys. Lett. B}, in press.
\item{[8]} R. Cecchini and E. Celeghini, {\sl Nuovo Cimento} {\bf
A37}, 266 (1977).
\item{[9]} J. Lukierski, H. Ruegg, W. R\"uhl, {\it From
$\kappa$--Poincar\'e algebra to $\kappa$--Lorentz Quasigroup:
a deformation of relativistic symmetry}, preprint KL-TH-92/22.
\item{[10]} L.C. Biedenharn, B. Mueller and M. Tarlini, {\it The Dirac-Coulomb
      problem for the $\kappa$--Poincar\'e Quantum Group}, preprint
Universit\`a di Firenze, March 1993.}

\vfill\break


\titolo{ Quantum algebras and basic hypergeometric functions }

\autori{\undertext{Roberto~Floreanini} and Luc Vinet}

The representation theory of quantum algebras is deeply
rooted in the theory of $q$-special functions.
These functions are extensions to a base $q$ of the standard special
functions. One of the best known example is the basic hypergeometric
series $(|q|<1)$ [1]
$$_2\phi_1(a,b;c;q,z)=\sum_{n=0}^\infty {(a;q)_n\, (b;q)_n\over
(q;q)_n\, (c;q)_n}\ z^n\ ,\qquad |z|<1\ ,$$
$$(a;q)_n=(1-a)(1-aq)\cdots(1-aq^{n-1})\ ,$$
which is the $q$-generalization of the usual hypergeometric series
$_2F_1$ of Gauss.
The following two $q$-exponentials are also very important:
$$\eqalign{&e_q(z)=\sum_{n=0}^\infty {1\over (q;q)_n}\, z^n\ ,
\qquad |z|<1\ ,\cr
           &E_q(z)=\sum_{n=0}^\infty {q^{n(n-1)/2}\over (q;q)_n}\,
z^n\ .}$$

The basic idea underlying the quantum algebra interpretation
of $q$-special functions is to replace the exponential mappings
from the Lie algebra $\cal G$ into the corresponding group by
$q$-exponentials from the quantum algebra ${\cal U}_q({\cal G})$
into the completion of it. The matrix elements of products of
$q$-exponentials in quantum algebra generators,
in specific representations, are then found to be
expressible in terms of $q$-special functions.

The hypergeometric function $_2\phi_1$ is in this way connected with
the quantum algebra ${\cal U}_q(sl(2))$, [5,6,7] while known
$q$-generalizations of classical orthogonal polynomials (Hermite,
Laguerre, Gegenbauer) appear in the representation theory of
$q$-oscillators and of ${\cal U}_q(su(1,1))$. [2,3,8]
The basic version of Bessel functions are instead
connected with the euclidean algebra
in two dimensions. [4,9]

This quantum algebra interpretation of $q$-special functions
is very useful since it allows deriving new properties
for these functions (orthogonality relations, generating functions,
addition formulas), hard to obtain otherwise.

As in the case of ordinary special functions,
the basic special functions enter the
solution of many physical problems: they arise
whenever quantum algebras are relevant to the description of
physical models ({\it e. g.} infinite spin chains and massive minimal models).
The study of the properties of these functions is thus
of great importance and the approach presented here seems to be
simple and promising.

\referenze{
\item{[1]} G. Gasper and M. Rahman, {\it Basic Hypergeometric Series},
(Cambridge University Press, Cambridge, 1990).
\item{[2]} R. Floreanini and L. Vinet, {\sl J. Math. Phys.} {\bf 33},
1358 (1992).
\item{[3]} R. Floreanini and L. Vinet, {\it Quantum algebras and
$q$-special functions}, {\sl Ann. of Phys.}, to appear.
\item{[4]} R. Floreanini and L. Vinet, {\sl J. Math. Phys.} {\bf 33},
2984 (1992).
\item{[5]} R. Floreanini and L. Vinet, {\it On the quantum group and
quantum algebra approach to $q$-special functions}, {\sl Lett. Math. Phys.},
to appear.
\item{[6]} R. Floreanini and L. Vinet, {\sl Phys. Lett.} {\bf A170}, 21 (1992).
\item{[7]} R. Floreanini and L. Vinet, {\it An algebraic interpretation
of the $q$-hypergeometric functions},
{\sl Journal of Group Theory in Physics}, to appear.
\item{[8]} E.G. Kalnins, H.L. Manocha and W. Miller, {\sl J. Math. Phys.}
 {\bf 33}, 2365 (1992).
\item{[9]} E.G. Kalnins, W. Miller and S. Mukherjee, {\it Models of
$q$-algebra representations: the group of plane motions},
University of Minnesota preprint, 1992.}

\vfill\break


\titolo{4--Dimensional lattice gravity and 12j-symbols}

\autori{Mauro Carfora, Maurizio Martellini and \undertext{Annalisa~Marzuoli}}

In 1968 Ponzano and Regge [1] discovered a deep connection
between the asymptotic expansion of a Racah-Wigner 6{\it j}-
symbol and the partition function for 3D-euclidean quantum
gravity with an action discretized according to the
prescription of {\it Regge Calculus} [2]. During the last
three years there have been important developments of this
original idea, which provide a new insight in the search for
the semiclassical limit of discretized 3D-gravity models and
which allow to discuss their connection with topological
quantum field theories (see [3]-[10]).\par

I present here a model (see [11],[12]) which provides a
4-dimensional version of Ponzano and Regge's result [1].
In particular, I show that the exponential of the euclidean
Einstein-Regge action for a 4D- discretized block is
given, in the semiclassical limit, by a suitable gaussian
integral of a 12{\it j}-symbol.
I also discuss a model of 4D-{\it topological} lattice
gravity which involves a 15{\it j}-symbol [13]. Differences
between the two models are stressed.

\referenze{
\item{[1]} G. Ponzano and T. Regge, in {\it Spectroscopic and
Group Theoretical Methods in Physics}, F. Block ed.
(North Holland, Amsterdam, 1968).
\item{[2]} T. Regge, {\sl Nuovo Cimento} {\bf 19}, 558 (1961).
\item{[3]} V. Turaev and O. Viro, {\sl Topology} {\bf 31}, 865 (1992).
\item{[4]} H. Ooguri and N. Sasakura, {\sl Mod. Phys. Lett.} {\bf A6},
3591 (1991).
\item{[5]} S. Mizoguchi and T. Tada, {\sl Phys. Rev. Lett.} {\bf 68},
1795 (1992).
\item{[6]} F. Archer and R.M. Williams, {\sl Phys. Lett.} {\bf B273},
438 (1991).
\item{[7]} H. Ooguri, {\sl Nucl. Phys.} {\bf B382}, 276 (1992).
\item{[8]} B. Durhuus, H.P.Jakobsen and R.Nest, {\sl Nucl. Phys.}{\bf B}
(Proc. Suppl.) {\bf 25A}, 109 (1992).
\item{[9]} M.Karowski, W. M\"uller and R. Schrader, {\sl Jour. of
Phys.} {\bf A25}, 4847 (1992).
\item{[10]} D.V. Boulatov {\it A model of 3-dimensional lattice
gravity}, Preprint SPhT/92-017 (1992).
\item{[11]} M.Carfora, M. Martellini and A. Marzuoli, {\sl Phys. Lett.}
{\bf B299}, 229 (1993).
\item{[12]} M. Carfora, M.Martellini and A. Marzuoli,{\it
4--Dimensional lattice gravity and \hfill\break 12j--symbols} Preprint
FNT/T 92/48 (1992).
\item{[13]} H. Ooguri, {\sl Mod. Phys. Lett.} {\bf A7}, 2799 (1992).}

\vfill\break


\titolo{The geometrical meaning of the quantum correction}

\autori{Davide Franco and \undertext{Cesare~Reina}}

The ring (Frobenius algebra) of local observables for
topological $\sigma$-models on the Riemann sphere $P^1$ with
values in the grasmanian $G(s;n)$ is known to be the
quotient of the cohomology ring of the target space by the
(inhomogeneous) ideal generated by the so-called quantum
correction. While the need for a quantum correction comes
from algebraic motivations in field theory, the aim of our
paper [1] is to understand its geometrical meaning.
The simple examples of $P^1 \rightarrow P^n$ models tell us
that the quantum correction is a form of Poincar\'e duality
which allows to compute intersections on moduli spaces of lower
degrees. We will check this point of view for the case
of $P^1\rightarrow G(s;n)$ models, yielding a proof of
the algebraic result from physics [2] in terms of the geometry of the
$\sigma$-model itself. Finally we generalize our
geometrical reasoning to flag-manifold valued models,
getting a new result which needs some physical
interpretation.

\referenze{
\item{[1]} D. Franco and C. Reina {\it The geometry of the quantum correction
for topological $\sigma$-models}, Preprint SISSA - FM/235/92
\item{[2]} K. Intriligator, {\it Mod. Phys. Lett.} {\bf A6}, 3543 (1991).}

\vfill\break

\centerline{\bf Participants}

\bigskip
\bigskip
\bigskip

\halign{\indent#\ &\quad#\hfil\cr

Francesco Bonechi,&Dipartimento di Fisica, INFN Sezione di Firenze\cr
                 &Largo E. Fermi 2, 50125 Firenze\cr
                 &Tel. 055-2298141, Fax. 055-229330\cr
                 &vaxfi::bonechi\quad\quad\quad bonechi@fi.infn.it\cr

Loriano Bonora,&SISSA and  INFN, Sezione di Trieste,
               via Beirut 2, Trieste,\cr
               &Tel. 040-3787436, Fax. 040-3787528\cr
               &38028::bonora \quad\quad\quad
               bonora@trieste.sissa.it\cr

Leonardo Castellani,&INFN Sezione di Torino, Dipartimento
               di Fisica Teorica,\cr
               &via P. Giuria 1, 10125 Torino,\cr
               &Tel. 011-6527206, Fax. 011-6699579\cr
               &vaxto::castellani\quad\quad\quad
               castellani@to.infn.it\cr

Enrico Celeghini,&Dipartimento di Fisica, INFN Sezione di Firenze\cr
                 &Largo E. Fermi 2, 50125 Firenze\cr
                 &Tel. 055-2298141, Fax. 055-229330\cr
                 &vaxfi::celeghini\quad\quad\quad celeghini@fi.infn.it\cr

Paolo Cotta-Ramusino,&Dipartimento di Fisica, MI\cr
                     &via Celoria 16, 20124 Milano\cr
                    &Tel. 02-2392425, Fax 02-2392480\cr
                    &vaxmi::cotta\quad\quad\quad cotta@mi.infn.it\cr

Ludwik Dabrowski,&ICTP, Strada Costiera 11, 34014 Trieste,\cr
                 &Tel. 040 3787422,\cr
                 &dabrowski@itssissa\cr

Gaetano Fiore,&SISSA, Via Beirut 2, 34014 Trieste,\cr
              &Tel. 040-3787434, Fax 040-3787528\cr
              &tsmi19::fiore\quad\quad\quad fiore@itssissa.it\cr

Roberto Floreanini,&INFN, c/o Dipartimento di Fisica Teorica,\cr
                   &Strada Costiera 11, 34014 Trieste,\cr
                   &Tel. 040-2240341, Fax. 040-224106\cr
                   &38472::florean\quad\quad\quad florean@trieste.infn.it\cr

Riccardo Giachetti,&Dipartimento di Matematica, Universit\`a di
                     Bologna\cr
                   &INFN Sezione di Firenze, Largo E. Fermi 2, 50125 Firenze\cr
                   &Tel. 055-2298141, Fax. 055-229330\cr
                   &vaxfi::giachetti\quad\quad\quad giachetti@fi.infn.it\cr}
\vfill\break
\halign{\indent#\ &\quad#\hfil\cr

Enore Guadagnini,&Dipartimento di Fisica, INFN Sezione di Pisa,\cr
                 &Piazza Torricelli 2, 56100 Pisa,\cr
                 &Tel. 050-45221, Fax. 050-48277\cr
          &mvxpi1::guadagnini\quad\quad\quad
            guadagnini@mvxpi1.difi.unipi.it\cr

Decio Levi,&Universit\'a di Roma III,
            INFN Sezione di Roma,\cr
           &P.le Aldo Moro 2, 00185 Roma,\cr
           &Tel. 06-49914284, Fax 06-4957697\cr
           &40221::levi\quad\quad
            levi@roma1.infn.it\quad\quad levi@sci.uniroma1.it\cr

Luca Lusanna,&INFN Sezione di Firenze, Dipartimento di Fisica,\cr
               &Largo E. Fermi 2, 50125 Firenze,\cr
               & Tel. 055-2298141, Fax. 055-229330\cr
               &vaxfi::lusanna\quad\quad\quad lusanna@fi.infn.it\cr

Luigi Martina,& Dipartimento di Fisica, Universit\`a degli Studi di Lecce\cr
              &via Arnesano, 73100 Lecce, Italia, CP 193,\cr
	      &Tel 0832-620446, Fax 0832-620505\cr
              &vaxle::martina\quad\quad\quad martina@lecce.infn.it\cr

Annalisa Marzuoli,&Dipartimento di Fisica Nucleare e Teorica, INFN
                    Sez. di Pavia,\cr
                  &via Bassi 6, 27100 Pavia,\cr
                  &Tel. 0382-392442, Fax 0382-526938\cr
                  &vaxpv::marzuoli\quad\quad\quad
                    marzuoli@pavia.infn.it\cr

Mihail Mintchev,&INFN Sezione di Pisa, Dipartimento di Fisica,\cr
                &Piazza Torricelli 2, 56100 Pisa,\cr
                &Tel. 050-501710, Fax. 050-48277\cr
                &mvxpi1::michele\cr

Arianna Montorsi,& Dipartimento di Fisica and Unit\'a INFM,
                   Politecnico di Torino,\cr
                 &C.so Duca degli Abruzzi 24, 10129 Torino\cr
                 &Tel. 011-5647318 Fax. 011-5647399\cr
                 &32202::montorsi\cr

Oktay K. Pashaev,&Dipartimento di Fisica, Universit\`a degli Studi di Lecce\cr
                 &via Arnesano, 73100 Lecce, Italia, CP 193,\cr
	         &Tel. 0832-620446, Fax 0832-620505\cr
                 &vaxle::pashaev\quad\quad\quad pashaev@lecce.infn.it\cr
                 &Joint Institute for Nuclear Research, 141980 Dubna, Russia\cr
                 &pashaev@main1.jinr.dubna.su\quad\quad
                   oktay@lcta.jinrc.dubna.su\cr

Orlando Ragnisco,&Universit\`a di Roma III,
                   INFN Sezione di Roma,\cr
                 &P.le Aldo Moro 2, 00185 Roma\cr
                 &Tel. 06-4451728, Fax  06-4957697\cr
                 &vaxrom::ragnisco\quad\quad\quad
                  ragnisco@roma1.infn.it\cr

Mario Rasetti,&Dipartimento di Fisica and Unit\'a INFM, Politecnico di
               Torino,\cr
              &C.so Duca degli Abruzzi 24, 10129 Torino\cr
              &Tel. 011-5647324, Fax. 011-5647399\cr
              &32144::rasetti\cr

Cesare Reina,&SISSA, Strada Costiera 11, 34100 Trieste,\cr
             &Tel. 040-3787441, Fax 040-3787528\cr
             &sissa::reina\quad\quad\quad reina@itssissa\cr

Antonio Sciarrino,&Dipartimento di Scienze Fisiche and INFN
              Sezione di Napoli,\cr
              &Mostra d'Oltremare Pad. 19, 80125 Napoli,\cr
              & Tel. 081-7253422, Fax 081-614508 \cr
              &vaxna1::sciarrino\quad\quad\quad
              sciarrino@vaxna1.na.infn.it\cr

Stefano Sciuto,&Dipartimento di Fisica Teorica and INFN Sezione di Torino,\cr
               &via P. Giuria 1, I-10125 Torino,\cr
               &Tel. 011-6527211, Fax 011-6527214\cr
               &31890::sciuto\quad\quad vaxto::sciuto\quad\quad
            sciuto@to.infn.it\cr

Giulio Soliani,&Dipartimento di Fisica, Universit\`a degli Studi di Lecce,\cr
           &via Arnesano, 73100 Lecce, Italia, CP 193,\cr
	   &Tel 0832-620446, Fax 0832-620505\cr
           &vaxle::soliani\quad\quad\quad
           soliani@lecce.infn.it\cr

Emanuele Sorace,&INFN Sezione di Firenze, Dipartimento di Fisica,\cr
               &Largo E. Fermi 2, 50125 Firenze,\cr
               & Tel. 055-2298141, Fax. 055-229330\cr
               &vaxfi::sorace\quad\quad\quad sorace@fi.infn.it\cr

Marco Tarlini,&INFN Sezione di Firenze, Dipartimento di Fisica,\cr
               &Largo E. Fermi 2, 50125 Firenze,\cr
               & Tel. 055-2298141, Fax. 055-229330\cr
               &vaxfi::tarlini\quad\quad\quad tarlini@fi.infn.it\cr

Piero Truini,&Dipartimento di Fisica e INFN Sezione di Genova,\cr
             &via Dodecaneso 33, 16146 Genova,\cr
             &Tel. 010-3536209, Fax 010-313358\cr
             &vaxge::truini\quad\quad\quad truini@genova.infn.it\cr

Giuseppe Vitiello,&Dipartimento di Fisica, Universita' di Salerno,
                  84100 Salerno\cr
                  &Tel. 089-822311, Fax. 089-953804\cr
                  &vaxsa::vitiello\quad\quad\quad
                     vitiello@sa.infn.it\cr

Francesco Zaccaria,&Dipartimento di Scienze Fisiche and INFN
              Sezione di Napoli,\cr
              &Mostra d'Oltremare Pad. 19, 80125 Napoli,\cr
              & Tel. 081-7253408, Fax 081-614508 \cr
              &vaxna1::zaccaria\quad\quad\quad
              zaccaria@vaxna1.na.infn.it\cr}

\bye